\documentclass[final,5p,authoryear]{elsarticle}
\usepackage{amssymb}
\usepackage{acronym}
\usepackage{stfloats}
\usepackage[table,xcdraw]{xcolor}
\usepackage{amsmath,amsfonts}

\journal{arXiv}

\begin{document}

\begin{frontmatter}

\title{Exhaust Gas Optimization of Modern Scooters by Velocity Control}
\author[1,2]{Jannis Kreß \corref{cor1}}
\ead{jannis.kress@fb2.fra-uas.de}
\author[1]{Jens Rau}
\author[1]{Ingo Behr}
\author[1]{Bernd Mohn} 
\author[1]{Hektor Hebert} 
\author[2]{Arturo Morgado-Estévez}

\cortext[cor1]{Corresponding author.}
\newcommand{\abbreviations}[1]{%
  \nonumnote{\textit{Abbreviations:\enspace}#1}}

\affiliation[1]{organization={Dep. of Computing and Engineering, Frankfurt University of Applied Sciences},
             city={Frankfurt},
             postcode={60318},
             state={Hessen},
             country={Germany}}
\affiliation[2]{organization={Dep. of Automation, Electronics and Computing Architecture and Networks, University of Cadiz},
             city={Puerto Real},
             postcode={11519},
             state={Andalusia},
             country={Spain}}

\begin{abstract}
This paper investigates the optimization of the exhaust gas composition by applying a velocity-controlled Throttle-by-Wire-System on modern 50~cc scooters (Euro~5). Nowadays combustion-powered scooters are still inefficiently restricted, resulting in an unreasonably high fuel consumption and unfavorable exhaust emissions. The velocity control prevents restriction by negatively shifting the ignition timing and regulates the throttle valve opening instead. Injection quantity, engine speed, ignition timing, cylinder wall temperature, exhaust gas temperature, oxygen sensor data, crankshaft position and in-cylinder pressure were acquired to measure engine parameters. At the same time, vehicle data on the CAN bus, such as throttle opening angle, the rider's acceleration command and vehicle velocity were recorded. For determination of the exhaust gas composition, five probes were sensing CO, CO$_2$, NO$_x$, O$_2$ and HC in addition to the temperature and mass flow. A \textit{Peugeot Kisbee 50 4T} (Euro 5) serves as test vehicle. The original and the optimized restriction were subjected to various gradients on a roller dynamometer at top speed. Thus, a statement can be made about all operating points of restriction. The resistance parameters required, were previously determined in a coast down test. When driving on level ground, a difference of 50\% in the throttle opening leads to a 17\% improvement in fuel economy. By measuring the engine parameters, optimum ignition timing could be proven with increasing internal cylinder pressure. Further, 17\% reduction in exhaust gas flow was demonstrated. CO emissions decreased by a factor of 8.4, CO$_2$ by 1.17 and HC by 2.1 while NO$_x$ increased by a factor of 3.
\end{abstract}


\end{frontmatter}

\newacro{CVT}{Continuously Variable Transmission}
\newacro{ECU}{Electronic Control Unit}
\newacro{HMI}{Human Machine Interface}
\newacro{I/O}{Input/Output}
\newacro{EC}{Engine Controller}
\newacro{EFM}{Exhaust Flow Meter}
\newacro{MB}{Measurement Box}
\newacro{MECU}{Main ECU}
\newacro{OR}{Original Restriction}
\newacro{TbWS}{Throttle-by-Wire-System}
\newacro{TDC}{Top Dead Center}
\newacro{TPS}{Throttle Position Sensor}
\newacro{TSS}{Transmission Speed Sensor}
\newacro{TVA}{Throttle Valve Actuator}
\newacro{VC}{Velocity Control}

\section{Introduction}
\begin{figure}[!h]
\centering
\includegraphics[width=8.8cm]{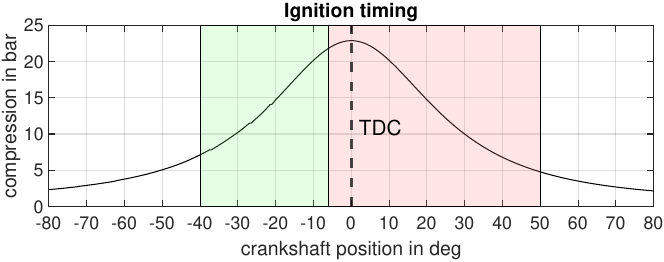}
\caption{Ignition timing \citep{VeloCtr_JK}}
\label{IgnT}
\end{figure}

The demand on individual transportation is increasing, while requirements on climate protection are being tightened by the European Climate Change Act \citep{EUcca}. Modern Euro~5 scooters could serve as an eco-friendly alternative to cars, reasoned by better fuel-economy and thus a minimized CO$_2$ footprint \citep{EURO5}. To meet the Euro~5 standard, today's scooters are equipped with a regulated catalytic converter and fuel injection. Legislation in the EU sets a max. speed of 45~km/h for this class of vehicle \citep{Def_scooter}. In the past, speed limitation was often realized through mechanical restrictors, manipulating the transmission ratio or mixture/exhaust flow \citep{R1}. Euro 2 certified scooters were frequently restricted by leaning, which led to high NO$_x$ emissions despite 3-way catalytic converters \citep{leaning}. Nowadays engine controllers are mostly shifting the ignition timing to limit the engine's performance. Reducing the injection quantity would lead to an unfavorable stoichiometric ratio ($\lambda > 1$) \citep{lambda} and thus negatively affect the functioning of the catalyst \citep{R2}. The \ac{OR} is bypassed and replaced by velocity-dependent control of the throttle valve position (air supply) \citep{VeloCtr_JK}. This is made possible by integrating a \ac{TbWS} \citep{TbWS_JK}, resulting in an optimum ignition timing while less fuel is injected to remain $\lambda = 1$. The ignition timing is dependent on engine speed and varies between 6° to 40° before the \ac{TDC}. For restriction, the timing is delayed and causes an inefficient engine operation. Consequently, combustion takes place during the downward movement of the piston (expansion) \citep{IgnT}. Figure \ref{IgnT} shows the optimum ignition timing left and the delayed timing right. The compression curve shown, corresponds to that of the test vehicle's engine. 

Suppressing the \ac{OR} and controlling the air inflow instead, leads to a drastic reduction of exhaust pollutants. To the best of our knowledge, this is the first application of a velocity-controlled \ac{TbWS} on a modern four-stroke 50~cc scooter for restriction purposes. The following novelties will make a substantial scientific contribution to the efficiency and environmental friendliness of scooters. 
\begin{itemize}
    \item The efficiency is increased and the injection quantity is reduced by 17\% at top speed.
    \item Limiting the air supply is directly minimizing the exhaust mass flow by 17\% while maintaining $\lambda = 1$. 
    \item Through an optimized combustion process, the exhaust gas composition is positively affected by lowered CO, CO$_2$ and HC emissions.
    \item Exhaust gas temperatures are lowered by up to 300°C, reducing thermal stress to the exhaust system.  
\end{itemize}

\subsection{Background on velocity-controlled \acp{TbWS}}
\acp{TbWS} are frequently used on high-performance motorcycles \citep{TbWS_SportMotorbike, TbWS_BMW1, TbWS_BMW2}. A \ac{TbWS} describes the replacement of a mechanical connection (throttle cable) with an electronic signal. The rider's throttle command is detected by a sensor and processed by an \ac{ECU}, which controls an actuator on the throttle valve. \acp{TbWS} has also been applied to small-volume scooters. The aim was better control in the event of interference \citep{EngineManagement} or more efficient operation of mechanically restricted two-stroke scooters \citep{ETC_Peugeot}.  A low-cost \ac{TbWS} was specifically developed for this research project based on a magnetoresistive throttle position sensor and a stepper motor-based throttle valve actuator. The system provides 99.63\% accuracy with a max. response time of 60~ms \citep{TbWS_JK}, enabling the influence of driver assistance systems. A \ac{VC} has been implemented as intelligent and eco-friendly restriction, leading to road-tested fuel savings of 14\% (\ac{OR}: 2.11~l/100~km, \ac{VC}: 1.82~l/100~km) while maintaining driving performance. A virtual dashboard, redundant wheel speed sensors \citep{WSS_JK} and a main \ac{ECU} have been developed. The \ac{VC} prevents restriction due to ignition timing shift and significantly improves combustion efficiency by controlling the throttle valve. An adaptive PI-controller is performing precise and stable vehicle velocity control \citep{VeloCtr_JK}.

\subsection{Background on combustion \& exhaust composition}
Four-stroke gasoline engines burn fuels consisting of multiple short-chain hydrocarbons, also called CHO compounds. A perfect and complete combustion would convert CHO to carbon dioxide (CO$_2$) and water (H$_2$O) through oxidation. During the combustion process, several reaction stages are passed through, temporarily producing hydrogen (H), H$_2$, oxygen (O), O$_2$, hydroxide (OH), carbon monoxide (CO), CO$_2$ and H$_2$O. Only at high temperatures, CO and H$_2$ are initially formed before CO oxidizes to CO$_2$ and H$_2$ to H$_2$O \citep{Combustion}. In terms of emissions, on the one hand unburned components such as CO and hydrocarbons (HC) on the other nitrogen oxides (NO$_x$) are critical constituents. The gases are produced as follows:

\begin{itemize}
    \item CO$_2$: A complete combustion of carbon converts a max. amount of CO$_2$ (if $\lambda$ = 1).
    \item CO: It is formed as an intermediate stage of carbon dioxide formation and in incomplete combustion under oxygen deficiency.
    \item HC: Unburned hydrocarbons are formed by the attachment of unreacted molecules to the cylinder wall, especially in the absence of oxygen.  
    \item NO$_x$: It is formed during combustion by nitrogen and oxygen. A surplus of air and high combustion temperatures lead to high emission.  
\end{itemize}

When the Euro~2 standard was introduced, the emission reduction of 50~cc scooters was investigated for different drive types. In particular, four-stroke engines with direct injection and catalytic converter were found to be promising. CO and HC emissions were significantly lower compared to carburetor-fueled engines or two-stroke engines \citep{Pot50cc}. To convert pollutants in a Euro~5 compliant manner, regulated 3-way catalytic converters are usually used with four-stroke engines and fuel injection. CO and HC oxide to CO$_2$ and H$_2$O, while reduction converts NO$_x$ to nitrogen (N) and CO$_2$ \citep{CAT}. The resulting conversion rates are better than 90\% as long as $\lambda = 1$ \citep{CATeff}. Accordingly, pollutants cannot be completely converted and the formation of the raw gases depends on the oxygen concentration of the mixture.

\section{System evaluation}
For a reproducible evaluation, the test runs are performed on a roller dynamometer. Parameter determination for dynamometer settings, measurement tools and evaluation strategy are described below. 

\subsection{Coast down test}
 To set suitable roller dynamometer parameters, the rolling and air resistances are determined by a coast down test, to ensure valid test conditions. These were carried out in calm wind conditions on a level and straight test track. First, the frontal area of the test vehicle, including rider (height: 1.8~m) and helmet, was visually determined to be 0.78~m$^2$. In order to compensate small differences in gradient, path-time curves of two test runs in opposite directions of travel were averaged. Based on the parameters shown in Table \ref{params_R} and the recorded path-time curves, a general driving resistance polynomial (\ref{eq.1}) is obtained for a \textit{Peugeot Kisbee 50 4T} (Euro~5). \newpage
 
\begin{table}[!h]
\caption{List of coast down test parameters}
\footnotesize
\label{params_R}
\begin{center}
\begin{tabular}{ c l c c c }
\hline
\textbf{}   & \textbf{Property}       & \textbf{Symbol}   & \textbf{Value}    \\ \hline
1           & Frontal area            & $A$               & 0.78 m$^2$        \\
2           & Tyre pressure           & $P$               & 2.3 bar           \\
3           & Moment of inertia factor& $T_in$            & 1.04              \\
4           & Mass scooter            & $m_{sc}$          & 99 kg             \\
5           & Mass rider              & $m_r$             & 80 kg             \\
6           & Air density             & $\rho$            & 1.232 kg/m$^3$    \\ \hline
\end{tabular}
\end{center}
\end{table}

\vspace{-0.5cm}

\begin{equation}
    F(v)= 0.015{v^{2}}+41.65
    \label{eq.1}
\end{equation}

Equation \ref{eq.1} is giving the needed force to overcome the velocity-dependent air resistance and constant rolling resistance at a certain velocity. The determined drag coefficient ($c_w$) is 0.7 and the rolling resistance coefficient ($f_R$) is 0.031. To verify the determined resistances, measurements of the throttle valve positions (TVP) at three reference velocities were taken. Therefore, the cruise control function of the velocity control was used. Afterwards the results were compared to throttle valve positions measured on the roller dynamometer. Table \ref{throttleCompare} proves the validity of the parameter set, since the throttle position is proportional to the required driving force.

\begin{table}[!h]
\caption{List of throttle valve positions}
\footnotesize
\label{throttleCompare}
\begin{center}
\begin{tabular}{c c c}
\hline
\textbf{Velocity}   & \textbf{TVP$_{Road}$}    & \textbf{TVP$_{Dyno}$}    \\ \hline
25 km/h             & 16 \%                         & 16 \%               \\
35 km/h             & 26 \%                         & 27 \%               \\
45 km/h             & 45 \%                         & 45 \%               \\ \hline
\end{tabular}
\end{center}
\end{table}

\vspace{-0.5cm}

\subsection{Vehicle sensor integration}
For evaluation, various quantities must be recorded to assess the effects on mixture preparation, engine and finally exhaust gas composition. This was realized by the development of a measurement box and the integration of multiple sensors.\\

\textbf{Measurement box}: The developed measurement box logs \ac{TbWS} data and senses spatial vehicle movement. In this use case, the command variable (set velocity), the current throttle position (manipulated variable) and vehicle speed (controlled variable) are recorded. In addition, a small circuit processes the signal from the inductive crankshaft position sensor and provides an engine speed signal. To be able to make a statement about the air-fuel mixture, the signal voltage of the lambda sensor is transferred to an oxygen concentration by means of a look-up table. Finally, the cyclic opening time of the injection valve is sensed and stored as a duty cycle.\\ 

\begin{figure}[!h]
\centering
\includegraphics[width=8.8cm]{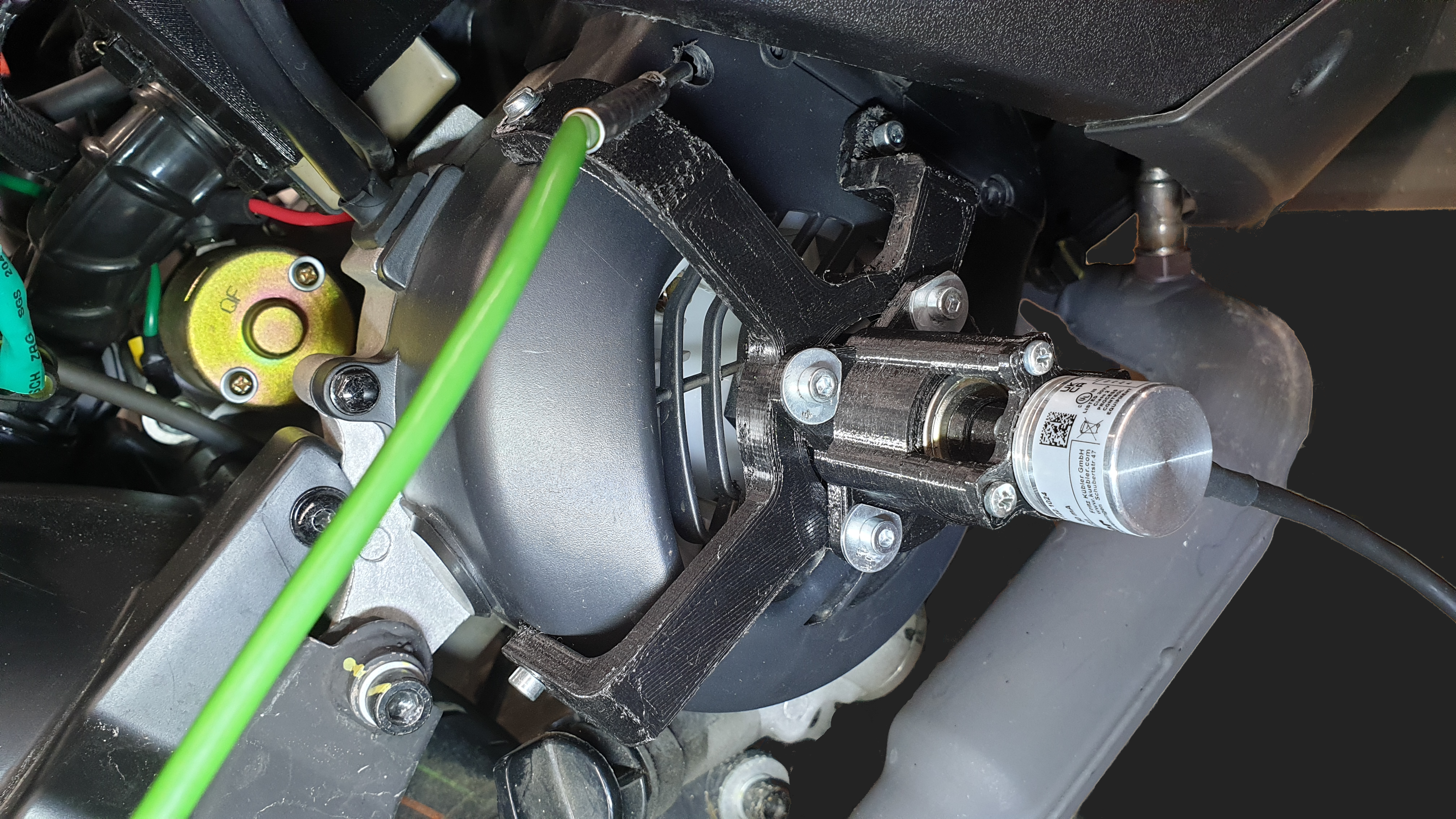}
\caption{Encoder adaption}
\label{EngAdap}
\end{figure}

\textbf{Encoder}: The original crankshaft sensor has a resolution of 24 steps, which makes an exact assignment of the measured engine parameters impossible. For this reason, a high-resolution and speed-resistant encoder must be adapted to the crankshaft. The decision was made for a \textit{Kübler 05.2400.1122.1024} encoder with a resolution of 1024 steps and a max. speed of 12000~rpm \citep{Encoder}. The motor speed reaches a maximum of 8000~rpm, but the small-volume single cylinder produces plenty of vibrations and therefore requires an adequate bearing/adaptation of the encoder. A torsionally rigid shaft-coupling combination made from a nylon carbon fiber (3D print) was placed on the fan wheel, which connects to the encoder. A dual bearing carrier made from PETG supports on the fan housing and allows alignment in radial and tumbling motion. The assembly is shown in Figure \ref{EngAdap}. 

\begin{figure}[!h]
\centering
\includegraphics[width=8.8cm]{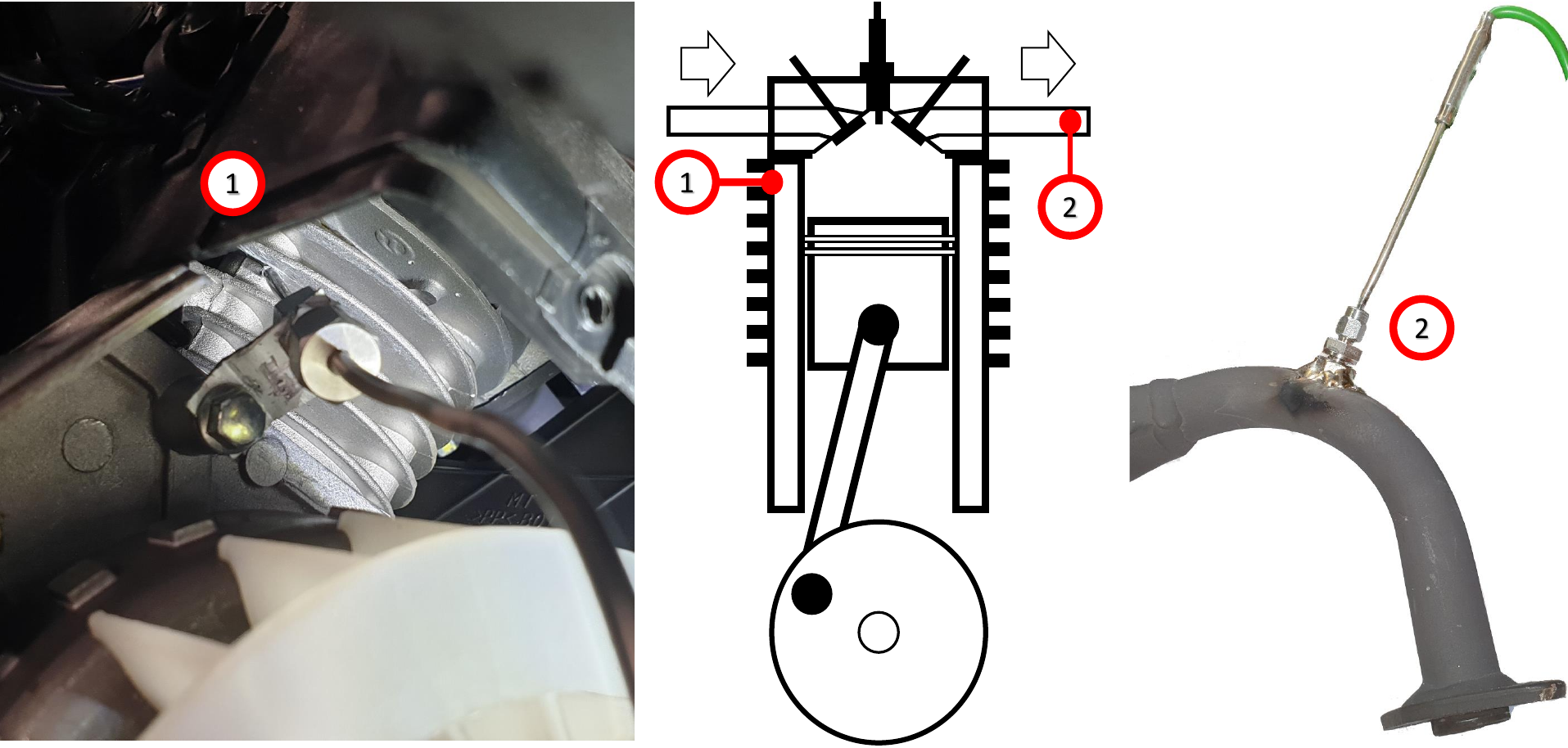}
\caption{Placement of thermocouples}
\label{ThermoPlacement}
\end{figure}

\textbf{Temperature probes}: Temperatures at the cylinder and in the exhaust manifold can allow conclusions to be drawn about combustion. Excess energy is converted into thermal energy in the \ac{OR} range. Thus, the temperature behavior depends on the ignition timing and the resulting combustion process. Two mineral-insulated thermocouples (Type K) are used, which have a temperature resistance of 1200°C \citep{Thermo}. Figure \ref{ThermoPlacement} shows the placements of both thermocouples. Probe 1 is positioned in a 2~mm deep hole in the cylinder wall and is guided past the fan wheel (left figure). Probe 2 has been welded directly into the exhaust manifold and is thus in the middle of the exhaust gas flow (right figure).  

\textbf{Indicating spark-plug}: Statements about the engine's operating cycle can only be made if ignition timing and in-cylinder pressure is measured. The ignition timing is tapped on the high-voltage ignition cable by using a measuring clamp. An indicating sparkplug (\textit{AVL ZI22 1\_U5D}) fitted with a piezo element is installed to measure the internal pressure \citep{Sparkplug}. Pressures of up to 200~bar can thus be measured without having to modify the combustion chamber. 

\subsection{Exhaust gas measurement}
Apart from the actual exhaust gas composition, the mass flow of the entire exhaust gas stream is of interest. For this purpose, so-called \acp{EFM} are used, which can determine the magnitude of the mass flow via a pressure difference. The selected \textit{AVL M.O.V.E} \ac{EFM} is equipped with a 2.5" exhaust pipe and has a measuring range of approx. 12.5-900~kg/h \citep{AVL_EFM}. The exhaust flow of a 50~cc four-stroke engine is very low despite high engine speeds due to the small displacement. Measurements on motorcycles have already been carried out using this method, but exhaust gases are usually collected for exhaust gas determination of such small engines. To estimate the mass flow of the test vehicle, a reverse calculation was performed based on the road tested fuel consumption ($Fuel_{con}$ = 1.00~l/h) \citep{VeloCtr_JK}. With \ac{OR} the expected exhaust mass flow ($m_{exh}$) at top speed is determined to 12~kg/h by Equation \ref{eq.2}. An investigation of measurements showed plausible mass flows at min. 7.9~kg/h. No valid measurements could be made at lower flow rates due to the inherent noise of the \ac{EFM}.

\begin{equation}
m_{exh}= (Fuel_{con}\cdot \rho_{fuel})\cdot (1 + 14.7\lambda)
\label{eq.2}
\end{equation}

The exhaust gas composition is determined with the portable emission measuring device \textit{AVL M.O.V.E Gas Pems}. The HC concentration is measured using the flame ionization detector principle. NO$_x$ is detected by a UV analyzer and CO/CO$_2$ by a non-dispersive infrared gas analyzer \citep{AVL_GP}. Before measuring, the system was calibrated and warmed up. Zero drifts were compensated periodically with zero gas.

\subsection{Data processing}
Engine related signals from the encoder, the ignition clamp and the measuring spark plug must be sampled at high rates. At max. engine speed, one crankshaft revolution takes approx. 8~ms. In order to use the full resolution of the crankshaft encoder (1024), the sampling rate would have to be at least 286~kHz according to the sampling theorem. The \textit{Sirius} high speed interface from \textit{ZSE Electronic} achieves a max. sampling rate of 1~MHz \citep{Sirius}. In order to compensate for fluctuating ignition times and mean pressures during measurement, 60~cycles are recorded and standardized afterwards. Data acquisition from the vehicle's internal CAN bus is carried at a sampling rate of 20~Hz. The processes are comparatively slow and static for the measurements to be carried out. The exhaust gas analysis generates standardized measurement data every second, based on a 10 Hz raw data sample rate.

\subsection{Test strategy}
\begin{figure}[!h]
\centering
\includegraphics[width=8.8cm]{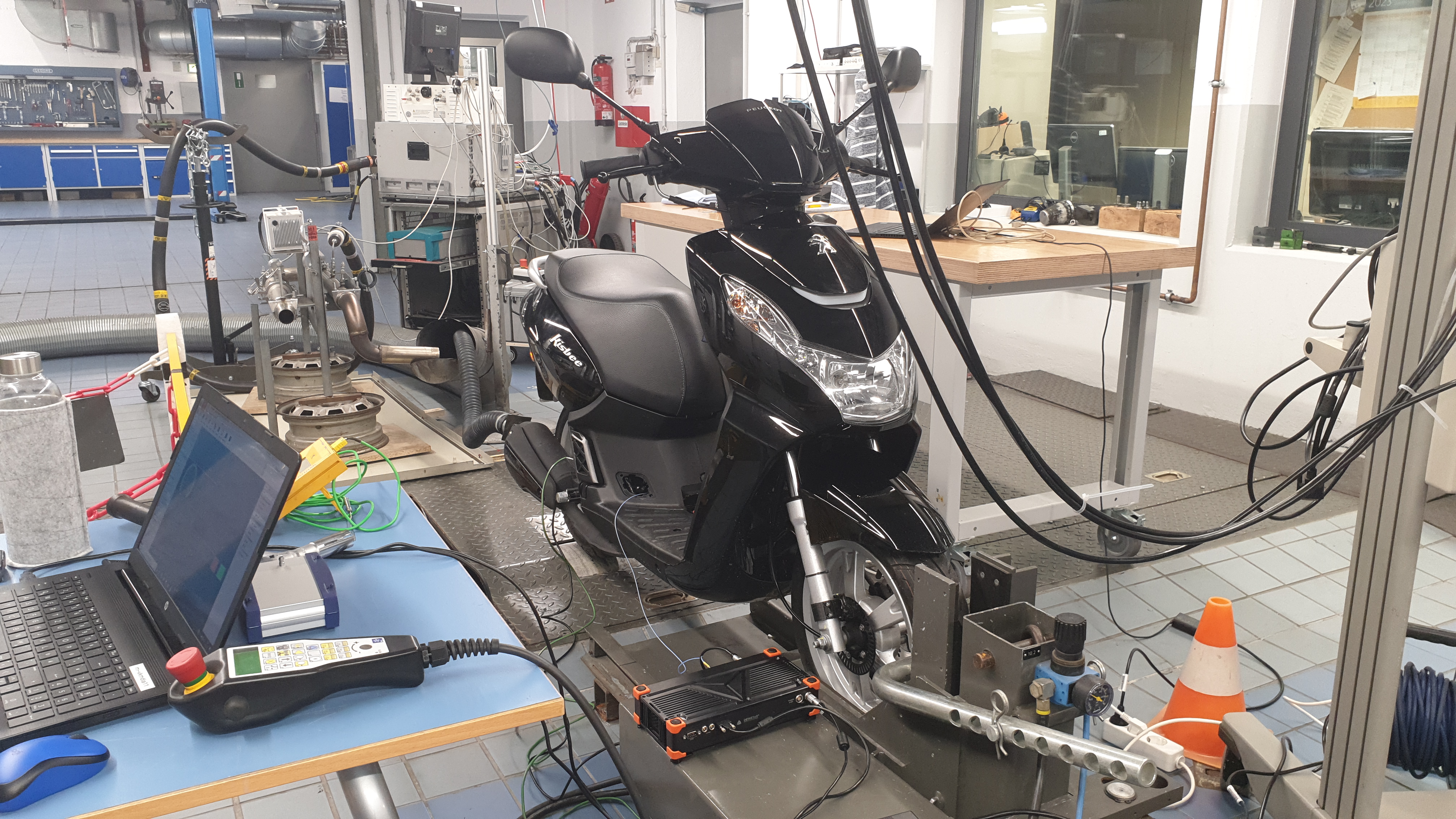}
\caption{Test setup}
\label{TestSetUp}
\end{figure}

Restriction by shifting the ignition timing only occurs when the max. permitted speed is reached. The measured top speed of the test vehicle (\ac{OR}) is 48.7~km/h in factory condition. In the remaining speed range, the driver manually adjusts the throttle valve position to achieve the optimum air supply. However, due to the low max. permitted speed of this vehicle class, it is usually driven at max. speed. For this reason, only operating points at 48.7~km/h are examined below. To obtain a load-dependent statement about the effect of the \ac{VC} compared to the \ac{OR}, a gradient is simulated using the roller dynamometer. Vehicle parameters and rider weight (80~kg) remain unchanged during all test runs, while the gradient is varied in 1\% steps in the range -8\% to 0\% and in 0.5\% steps between 0\% and 2\%. With a gradient of -8\%, the downhill force and the driving force are balanced. Due to the low engine power, the top speed can no longer be reached on gradients exceeding 2\%, which also prevents ignition timing manipulation. In order to obtain valid, reproducible and comparable measurements, the specified operating points are approached statically. The measurements are started once the top speed has been reached with the corresponding load (gradients) and the velocity control has been settled. Figure \ref{TestSetUp} shows the final setup on the roller dynamometer.  

\vspace{-0.4cm}

\section{Results}
The results are divided into vehicle CAN data, engine and exhaust gas according to the measurement data acquisition. Contrary to the test strategy described, only gradients down to -5\% could be taken into account. The mass flow could not be plausibilized for steeper gradients with activated \ac{VC} due to its small exhaust gas volume. Despite clustering, all the measurement data shown, relates to the same measurement run and was recorded simultaneously.

\subsection{Vehicle CAN bus data}
\begin{figure}[!h]
\centering
\includegraphics[width=8.8cm]{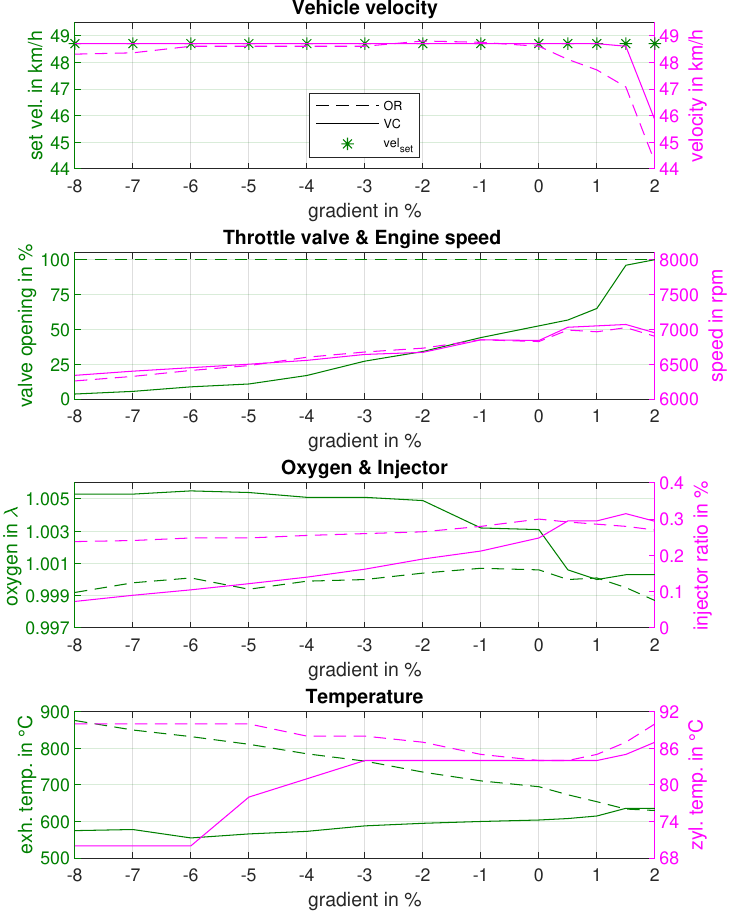}
\caption{CAN bus data}
\label{VeM}
\end{figure}

Figure \ref{VeM} shows the logged vehicle data recorded by the measurement box. The \textbf{Vehicle velocity} of the \ac{OR} and \ac{VC} differs in behavior. The \ac{VC} adjusts the top speed precisely until the engine power is no longer sufficient on gradients overcoming 1.5\%. With the \ac{OR}, the speed fluctuates slightly, with clear deviations on downhill gradients greater than 6\% (0.4~km/h) and uphill gradients (up to 1.6~km/h). Further, the \textbf{Throttle valve} behavior is demonstrated. Contrary to the original throttle cable actuation (fully open at top speed), the system controls the throttle valve position load-dependent instead of the ignition timing. A difference of 50\% can already be observed on a level road. In addition, the \textbf{Engine speed} shows the effects of changing load conditions. Small deviations depend on the differences in vehicle velocity caused by inaccurate \ac{OR}. In order to make a statement about the fuel-air mixture with regard to the ignition timing shift, the \textbf{Oxygen} sensor's measurement data was recorded. The oxygen concentration increases minimally with decreasing road gradient for \ac{VC} operation. In accordance to the stoichiometric ratio ($\lambda \approx$~1), the \textbf{Injector} opening must also be adjusted in proportion to the lowered air supply.

During activated \ac{VC}, the injection quantity increases in proportion to the engine load and throttle valve opening. With \ac{OR}, the injection quantity also increases slightly until a gradient of 0 \%. When driving uphill, the injection quantity decreases in line with the decreasing velocity, although the powertrain still offers power reserves. On level ground, the \ac{VC} achieved an improvement of 17\%. The \textbf{Temperatures} of the \ac{VC} show a proportional relation to the injection quantity, which in turn, depends on the throttle valve position. The cylinder temperature only differs on steeper gradients and improves by a max. of 22\% ($\Delta$20°C). The exhaust gas temperature is reduced by max. 34\% ($\Delta$301°C), thus drastically minimizing the thermal load on the exhaust system, even on level ground. Despite the reduction, the temperature remains within the optimum operating range for three-way catalytic converters. The temperature reduction can be explained by more efficient combustion. Instead of burning excess fuel in the exhaust gas, it is not injected at all. Tests of the engine parameters will confirm this in the following. 

\subsection{Effect on engine operation}
\begin{figure}[!h]
\centering
\includegraphics[width=8.8cm]{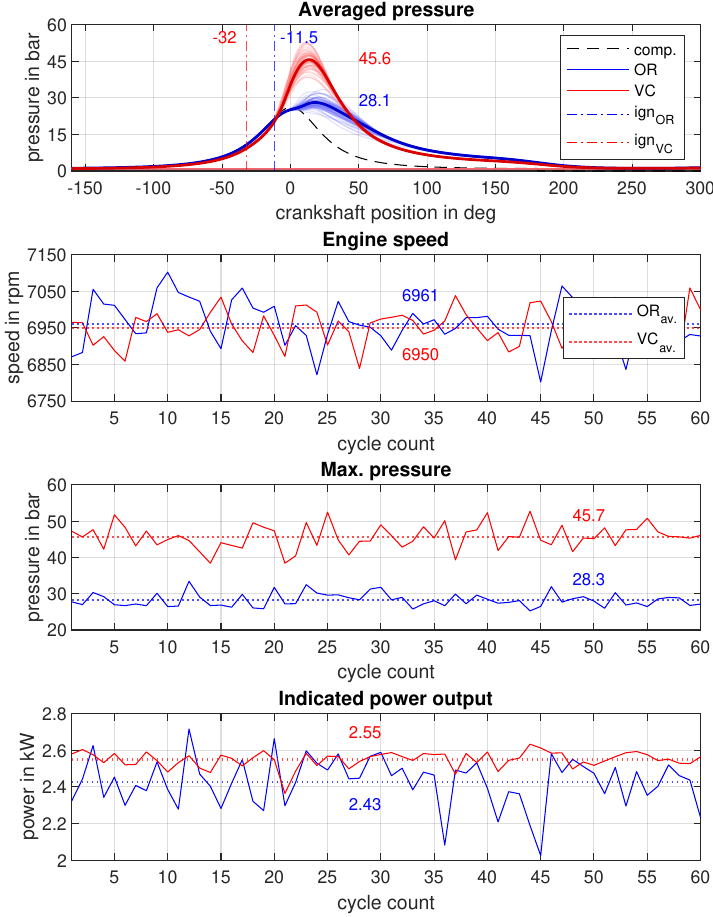}
\caption{Single engine operating point (0\%)}
\label{EngM}
\end{figure}

A measurement of the engine parameters was made for each operating point. As parameters vary slightly with each operating cycle due to the control and combustion processes, a measurement consists of 60~cycles. To illustrate the measurement methodology, an entire series of measurements for a gradient of 0\% is shown in Figure \ref{EngM}. All other operating points are evaluated summarized.

Differences in the indicated cylinder \textbf{Pressure} can be seen clearly. Forces acting on the piston, result directly from the cylinder pressure. Accordingly, the amount and duration of the existing pressure is decisive. The max. \textbf{Averaged pressure} shows an 63\% increase and the peak is slightly brought forward by \ac{VC}, which can be explained by an optimum ignition timing. Even on a level ground, the ignition timing is shifted forward by 20.5~degrees. The \textbf{Engine speed} does not differ noticeably, as the vehicle velocity is the same in both cases (see Fig.~\ref{VeM}). The \textbf{Max. pressure} measured differs minimally from the max. averaged pressure, which can be justified by the averaging. An \textbf{Indicated power output} can be calculated by the measured mean pressure and engine speed based on the engine's displacement. Sustained and increased mean pressures result in an improved power output.\\

\begin{figure}[!h]
\centering
\includegraphics[width=8.8cm]{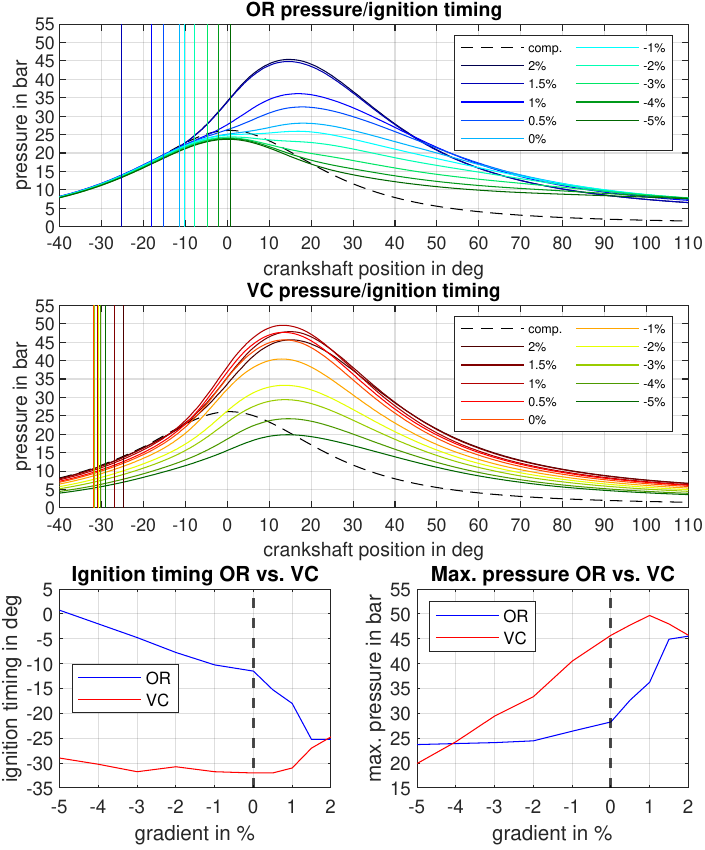}
\caption{Engine measurement}
\label{EngM2}
\end{figure}

Figure \ref{EngM2} shows the results of all gradient variations. The pressure curves of the \ac{VC} are more compressed and higher, while the \ac{OR} pressure drops slower. Consequently, the combustion process is much faster when \ac{VC} is enabled. Comparing the pressures only at max. load (2\%), the curves are almost identical in magnitude and progression, even if the vehicle velocity comes with a deviation (1.6~km/h). Major differences can be seen in the ignition timings. While the \ac{OR} shifts the ignition timing towards or slightly after the \ac{TDC}, the engine controller initiates the mixture ignition up to 29.75~degrees earlier with \ac{VC}.

According to the injection quantity in Figure \ref{VeM}, a distinction must be made between the gradient-dependent pressure curves. All pressures of the \ac{OR} result from the combustion of the max. injected fuel quantity. By shifting the pressure maximum (combustion) to expansion (downward movement of the piston), the energy is converted less effectively. The \ac{VC}, in contrast, achieves higher pressures using considerably less fuel. The ignition timing and max. pressure graphs illustrate the effect for driving on level ground. The steeper the gradient with \ac{OR}, the less of the energy provided is required and the more energy is burned in the exhaust. When \ac{VC} is activated, the ignition timing remains almost constant and is only adapted to the engine speed (see Fig.~\ref{VeM}). Even on level ground, the timing differs by 20.5~degrees and the mean pressure by 17.4~bar. 

\subsection{Exhaust optimization}
\begin{figure}[!h]
\centering
\includegraphics[width=8.5cm]{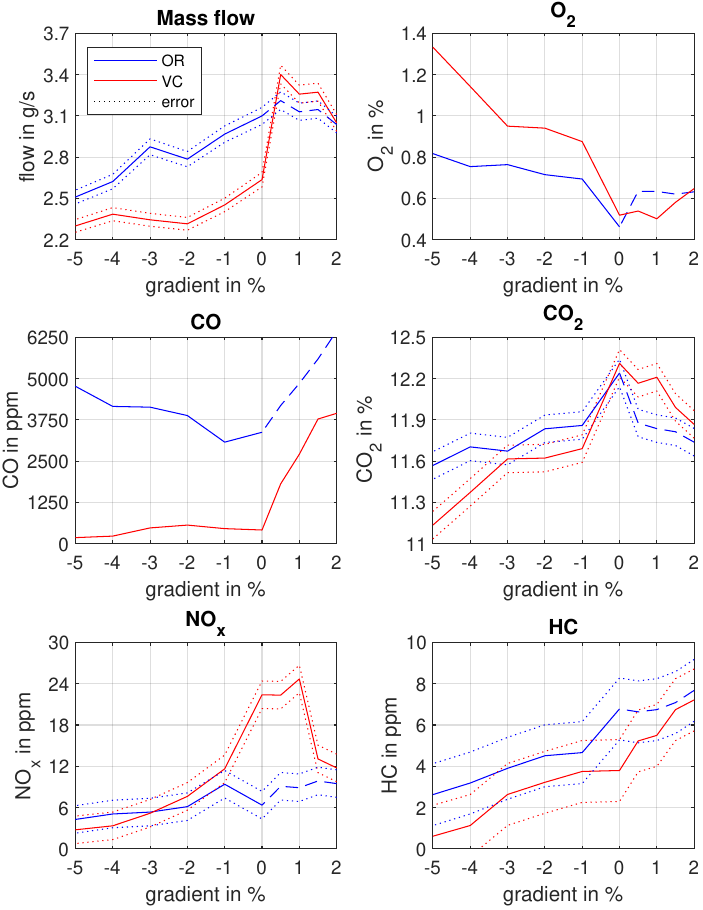}
\caption{Exhaust measurement}
\label{ExM}
\end{figure}

In addition to performance, all the engine parameters described have an influence on the exhaust composition. Figure \ref{ExM} shows the results of the exhaust gas measurement. 

Relative results are directly related to the \textbf{Mass flow} and mass flow improvements reduce the emitted emissions accordingly. Despite identical conditions for the measurement of both restriction methods, the results are only partially comparable. When evaluating the recorded vehicle velocity (see Fig.~\ref{VeM}), a deviation was noticed when driving uphill. This is due to a reduced injection quantity with \ac{OR}, although the engine would be able to maintain higher vehicle velocities. As a result, the mass flow is reduced, leading to different operating points (dashed line) that are not qualitatively comparable. No changes were made to the injection control as part of the \ac{VC} implementation. The given behavior can already be seen in the original state. The mass flow composes of the aspirated air and the injected fuel. When using the \ac{VC}, the throttle valve is not permanently open, less air is drawn in and less fuel is injected. Consequently, the exhaust mass flow is reduced. The optimized combustion within the cylinder (see Fig.~\ref{EngM2}) reduces the combustion in the exhaust, which is caused by the overlapping of the already opened exhaust valve. Only minor changes can be seen in the \textbf{CO$_2$} concentration. \\

The stoichiometric ratio and the combustion temperature have a major influence on the exhaust gas composition. Since only small variations of the oxygen concentration ($\Delta \lambda 0.007$) were detected during operation of both systems (see Fig.~\ref{VeM}), it can be assumed that stoichiometric changes will only have minor influence on the engine's raw exhaust gases. The catalytic converter, by contrast, reacts more sensitively to changes of the oxygen concentration (optimum: $0.99<\lambda <1.0$). For lean mixture, the catalyst's conversion rate of nitrogen oxides decreases in particular. Due to the faster combustion (see Fig.~\ref{EngM2}), the combustion temperature increases significantly, as the same amount of energy is converted in a shorter time. As a result, decreasing CO/HC and increasing NO$_x$ emissions are to be expected. The measurement actually shows extreme relative improvements in \textbf{CO} emissions by a max. factor of 26. The \textbf{HC} concentration also falls by a max. factor of 1.8. However, the increased temperatures favor the formation of NO$_x$ emissions. \textbf{NO$_x$} emissions increase under load by a max. factor of 3.5. The slight increase in \textbf{O$_2$} concentration can be attributed to the minimally increased lambda value and the catalytic reduction of NO$_x$. \\

If the \ac{OR} would exploit the powertrain's performance potential as the load increases, the exhaust gas emissions would equal those of the \ac{VC} when driving uphill. At an incline of 2\%, both operating points would be identical, as no restriction would limit the vehicle's velocity. The mass flow would grow due to the higher injection quantity. CO and HC emissions would decrease with rising temperature and the NO$_x$ emission would increase accordingly. When driving downhill, all emissions fall drastically below those of \ac{OR} operation. This would demonstrate emission improvements over all investigated load points. 50~cc scooters are mainly used in urban areas/on flat terrain. Table \ref{params} shows the relative improvements of the exhaust gas composition when driving on level ground. The resulting emissions are discussed subsequently.

\begin{table}[!h]
\caption{List of exhaust improvements on level ground}
\footnotesize
\label{params}
\begin{center}
\begin{tabular}{ c l c c c c c}
\hline
  &\textbf{Property} &\textbf{\ac{OR}} &\textbf{VC} &\textbf{Improvement}          \\ \hline  
1 &Mass flow in g/s  &3.1              &2.64        &\cellcolor[HTML]{32CB00}1.17  \\
2 &CO in ppm         &3373             &412.7       &\cellcolor[HTML]{32CB00}8.17  \\
3 &CO$_2$ in \%      &12.24            &12.3        &\cellcolor[HTML]{32CB00}1.00  \\
4 &NO$_x$ in ppm     &6.35             &22.4        &\cellcolor[HTML]{FE0000}-3.53 \\
6 &HC in ppm         &6.8              &3.8         &\cellcolor[HTML]{32CB00}1.79  \\ \hline 
\end{tabular}
\end{center}
\end{table}

\section{Discussion}
Classification of the absolute emission values is not directly feasible on the basis of the European emission standards. Here, the effects of the \ac{VC} were compared in relation to the \ac{OR}. Emission limits refer to measurements from standardized driving cycles (e.g. ECE R47, WMTC). Comparing emissions at a certain operating point can not be made with official emission data of other scooters. In addition, the WMCT cycle includes numerous accelerations and braking maneuvers between which the top speed is not reached \citep{WMTC}. During testing, the scooter predominantly reached the top speed, even in inner cities with high traffic density. Consequently, the exhaust measurements shown have a very clear effect in real-life use. The effect is maximized in non-urban use.\\   

In order to place the measured emissions in the context of the European emission standard, the measurements were converted approximately. For the following evaluation, the measurement is assessed on level ground. To allow the measured emissions to be classified, a conversion to mass per kilometer driven (g/km) must be performed. As the measurements describe volumetric proportions, the exhaust gas volume must be approximated by use of the general gas equation. As pollutants (CO, HC, NO$_x$) only represent 1\% of the exhaust gas, variations in relation to the total volume flow can be neglected. Its main components N$_2$ (71\%), CO$_2$ (14\%) and H$_2$O (13\%) remain mainly constant \citep{ExEm}. Further, ambient pressure (101325~Pa) and gas temperature (see Fig.~\ref{ExM}) must be taken into account. By inserting the parameters previously measured, an exhaust gas volume ($V_{Exh}$) of 0.321~m$^3$/min when using the \ac{OR} and 0.248~m$^3$/min with \ac{VC} results. Based on the volumetric pollutant proportions, the respective mass flow rate can be calculated using the general gas equation. By taking into account the top speed, the pollutant's volumes can be specified as masses in relation to a driven distance of one kilometer. Table \ref{disc} shows the converted emissions.

\begin{table}[!h]
\caption{List of emissions}
\footnotesize
\label{disc}
\begin{center}
\begin{tabular}{l c c c}
\hline
\textbf{Param.} & \textbf{\ac{OR}}  & \textbf{\ac{VC}}  & \textbf{Euro 5} \\ \hline
$V_{Exh}$       & 15.63 m$^3$/h     & 12.09 m$^3$/h     & /               \\
$V_d$           & 0.321 m$^3$/km    & 0.248 m$^3$/km    & /               \\ \hline
CO$_2$          & 43.1 g/km         & 36.81 g/km        & /               \\
CO              & 642.14 mg/km      & 76.48 mg/km       & 1000 mg/km      \\
HC              & 8.06 mg/km        & 3.84 mg/km        & 100 mg/km       \\
NO$_x$          & 2.33 mg/km        & 7.04 mg/km        & 60 mg/km        \\ \hline
\end{tabular}
\end{center}
\end{table}

The comparison with the Euro~5 emissions standard (valid from 2020) demonstrates that although the vehicle can comply with all limit values for this operating point, the \ac{OR} CO emissions in particular are relatively high. By using the \ac{VC}, CO emissions can be reduced by a factor of 8.4. As CO is a dangerous respiratory toxin for humans, the improvements are crucial. HC emissions are also reduced by a factor of 2.1, although these are far below the limit. Despite the 3-fold increase in NO$_x$ emissions, they are still low compared to the Euro~5 limit. CO$_2$ emissions are not taken into account by the exhaust gas standards, even though they are largely responsible for the greenhouse effect. An improvement by a factor of 1.17 can be observed. For gradients from -1\% to -8\%, the emissions would further improve significantly. For gradients $>$ 0\%, the emissions of both systems would equalize with increasing gradient.  

\section{Conclusion}
A velocity-controlled Throttle-by-Wire-System was developed in the preliminary stages, which serves as an eco-friendly restriction for 50~cc scooters and significantly reduces fuel economy. In order to evaluate the influence on the mixture preparation, the combustion process and exhaust gas formation, the vehicle was equipped with measurement technology. A measurement box logged all system-relevant CAN data. The engine was fitted with a crankshaft encoder, an indicating spark plug for pressure indication and an ignition clamp. Cylinder and exhaust gas temperature were measured by attaching two temperature probes. The exhaust gas composition was examined with regard to mass flow, CO, CO$_2$, NO$_x$, O$_2$ and HC concentration.\\

A coast down test was performed to determine the vehicle's resistance parameters. These were used to approach load points by varying the road gradient on a roller dynamometer at top speed. A considerable improvement in exhaust gas emissions was demonstrated caused by the velocity-controlled Throttle-by-Wire-System. The exhaust mass flow was reduced across all operating points (17\% on the level). Toxic CO emissions were also reduced by a factor of 8.4. Clear reductions (max. 54\%) in carcinogenic HC emissions were detected over the entire measurement range. NO$_x$ emissions increased for gradients between -1\% and 1\% by a max. factor of 3. CO$_2$ emissions, which are largely responsible for the greenhouse effect, were minimized by 17\%. These improvements are achieved, as demonstrated by measuring the engine parameters, by suppressing the ignition timing shift and regulating the air supply. The measurements show a max. ignition offset of 29.5~degrees between original restriction and velocity control. The internal cylinder pressures increase with velocity control operation and combustion is more efficient and faster. As a result, the combustion temperatures increase, which leads to the observed optimization of CO and HC emissions.

\section*{Acknowledgements}
The herein presented research was funded and supported by the University of Cadiz and Frankfurt University of Applied Sciences. The authors acknowledge the valuable support of Peugeot Motocycles Deutschland GmbH.


\bibliographystyle{elsarticle-harv} 
\bibliography{reference}



\end{document}